\documentclass[a4,usenatbib]{mn2e}

\usepackage{epsfig}
\usepackage{amsmath}
\usepackage{natbib}

\def\apj{ApJ}

\def\apjs{ApJS}
\def\mnras{MNRAS}

\def\nat{Nature}
\def\prd{Phys.Rev D}

\begin{document}

\title[Cosmological Constraints from SHAM analysis of galaxy clustering]{Cosmological Constraints from applying SHAM to rescaled cosmological simulations}
\author[Simha \& Cole]
{Vimal Simha$^{1}$, Shaun Cole$^{1}$\\
$^1$ Institute for Computational Cosmology, Department of Physics, Durham University, South Road, Durham DH1 3LE,\\
e-mail: vimalsimha@durham.ac.uk\\
} 

\maketitle


\begin{abstract}

We place constraints on the matter density of the Universe and the amplitude of clustering using measurements of the galaxy two-point correlation function from the Sloan Digital Sky Survey (SDSS). We generate model predictions for different cosmologies by populating rescaled N-body simulations with galaxies using the subhalo abundance matching (SHAM) technique. We find $\Omega_{\rm M}$ = 0.29 $\pm$0.03 and $\sigma_8$ = 0.86 $\pm$ 0.04 at 68\% confidence from fitting the observed two-point galaxy correlation function of galaxies brighter than $M_r$ = -18 in a volume limited sample of galaxies obtained by the SDSS. We discuss and quantify potential sources of systematic error, and conclude that while there is scope for improving its robustness, the technique presented in this paper provides a powerful low redshift constraint on the cosmological parameters that is complementary to other commonly used methods.

\end{abstract}


\newpage
\section{Introduction}

Determining cosmological parameters such as the matter density of the
Universe and the amplitude of clustering is a key goal of
cosmology. Measurements of the clustering of galaxies provide a unique
window into the distribution of dark matter in the Universe from which
cosmological parameters can be inferred. Galaxy surveys map the
distribution of visible baryons which indirectly trace the underlying
distribution of the gravitationally dominant dark matter. While the
distribution of dark matter for a given cosmology can be reliably
computed from first principles using cosmological N-body simulations,
because of uncertainties in the detailed physics of galaxy formation -
gas cooling, star formation, feedback etc., the distribution of
galaxies cannot be robustly predicted from first
principles. Consequently, one of the principal impediments in
inferring cosmological information from observations of galaxy
clustering is galaxy bias, the difference between the distribution of
galaxies and the underlying dark matter. In this paper, we use a
simple non-parametric model to relate observed galaxy luminosity to
halo mass in an N-body simulation, and use this model to place
constraints on the universal matter density ($\Omega_{\rm M}$) and the
amplitude of clustering ($\sigma_8$) by fitting the observed
clustering of galaxies.

We relate (sub)halos in our N-body simulation to observed galaxies
using subhalo abundance matching (SHAM). SHAM is a simple,
non-parametric model based on assuming a monotonic relationship
between observed galaxy luminosity and simulated (sub)halo mass. In this model, all galaxies are assumed to be contained within dark
matter subhalos, and galaxy luminosity is assumed to be monotonically
related to the present day subhalo mass for central subhalos and the
subhalo mass at the accretion epoch for satellite subhalos. SHAM takes only the space density of galaxies as input and predicts the 
clustering of a galaxy population. We use it to predict the clustering of 
observed volume limited samples of galaxies. Distinct cosmologies produce distinct populations of dark matter halos
and subhalos \citep{zheng02}, thus for each cosmology, the SHAM model
predicts a distinct galaxy auto correlation function.

SHAM  has been successfully used to match theoretical predictions to observables. For example, 
\cite{guo10} match the observed stellar mass - halo mass relation by populating N-body simulations
using the SHAM technique. \cite{tg10} match the observed circular velocity statistics using a similar method. \cite{simha12} found reasonable agreement between its assumptions and the 
output from a cosmological smoothed particle hydrodynamics (SPH) simulation
which incorporated gas physics, star formation and feedback from supernovae driven winds.

Running a suite of high resolution cosmological N-body simulations to
adequately sample the cosmological parameter space is computationally
demanding. Instead, we rescale the masses, positions and velocities of 
 halos/subhalos obtained from one simulation to different target cosmologies.
We use the reduced version of the rescaling technique
of \cite{angulo10} as implemented by \cite {ruiz11} to construct
subhalo catalogues for a given set of cosmological
parameters. \cite{angulo10,ruiz11,guo12} have shown that the relevant
properties of (sub)halos in such scaled models are very close to
those in simulations carried out directly with the target cosmology.

Our approach is similar in spirit to previous studies such as \cite{tinker11} 
who use halo occupation distribution (HOD) models, and \cite{harker07} who use semi-analytic models
and a similar simulation rescaling technique to fit the observed clustering
of galaxies to infer cosmological parameters. In contrast to the semi-analytic models of galaxy formation used by \cite{harker07}, we do not make any assumptions about 
the detailed gas physics that is involved in the process of galaxy formation. Although HOD models such as those used by \cite{tinker11} are also based on statistical descriptions rather than detailed modelling of the gas physics, there are significant differences between our techniques. HOD models describe galaxy bias using a probability
distribution P(N $\vert$ M$_h$), the probability that a halo of mass M$_h$ contains N
galaxies of a given type. In HOD models, the satellite occupation function
is constructed using a parametric model. In contrast, in SHAM, the satellite 
occupation function is determined by the N-body simulation. Consequently, HOD models have
several free parameters while the SHAM model used in this paper has none.

In \S2, we describe our methods - data, simulation, subhalo abundance matching and rescaling technique. In \S3, we
describe our theoretical model and how it is sensitive to different
cosmologies. In \S4, we discuss our results and constraints on cosmological parameters. In \S5, we discuss systematic uncertainties in our technique. In \S6, we discuss the utility of our model in fitting other data sets to obtain complementary
constraints on the cosmology. Finally, in \S7, we summarise our results.

\section{Methods}

\subsection{Data}

Our galaxy clustering measurements are obtained from 7900 deg$^2$ of sky 
observed by SDSS \citep{zehavi10}. We restrict our analysis to volume-limited
samples of galaxies

The clustering quantity we use for each sample is the projected auto
correlation function $w_p(r_p)$ defined as:
\begin{equation}
w_{\rm p}(r_{\rm p}) = 2 \int_0^{\pi_{\rm max}} \xi(r_{\rm p},\pi)d\pi,
\end {equation}

where $r_p$ is the projected separation between two galaxies, $\pi$ is the 
line-of-sight separation between two galaxies, and $\xi(r_p,\pi)$ is the 
measured two-dimensional correlation function. Due to the finite volume of the
survey, the integral is limited to $\pi_{\rm max}$ = 40 $h^{-1}$Mpc.

Our analysis is primarily focussed on a volume-limited sample of galaxies brighter than 
$M_r$ = -18.0. In addition, we make use of other volume-limited samples to test the robustness of our conclusions. The error estimates for each clustering sample are computed using the jackknife technique 
(see \citealt{zehavi10} for more details).

\subsection{Simulations}

We use two simulations, the Millennium Simulation \citep{springel05} and the higher resolution Millennium II 
simulation \citep{bk09}. Both simulations follow the evolution of 2160$^3$ particles from 
$z=127$ to $z=0$ in a $\Lambda$CDM cosmology (inflationary, cold dark matter with a 
cosmological constant) with $\Omega_{\rm M}$=0.25, $\Omega_{\Lambda}$=0.75,
$h\equiv H_0/100$ km s$^{-1}$Mpc$^{-1}$= 0.73, 
primordial spectral index $n_s$=1, and the amplitude of mass fluctuations , $\sigma_8$=0.9 where $\sigma_8$ is 
 the linear theory rms mass fluctuation amplitude in spheres of radius 8 $h^{-1}$Mpc at $z=0$. These parameter values 
 were chosen to agree with WMAP1 data \cite{spergel03}, and are 
different from but reasonably close to current estimates from the cosmic microwave background
\citep{larson10} and large scale structure \citep{reid10}, the main difference is that the more recent data favour a lower 
value of $\sigma_8$.

The Millennium Simulation (MS) simulates a comoving box that is 500$h^{-1}$Mpc on each side while the Millennium II Simulation (MS-II) simulates a comoving box that is 100$h^{-1}$Mpc on each side. The simulation particle masses are $m_p$ = 8.6$\times$10$^8$ $h^{-1}M_{\odot}$ in MS and 6.9$\times$10$^6$ $h^{-1}M_{\odot}$ in MS-II. 

For each output epoch in each simulation, the friends-of-friends (FOF) algorithm is used to identify groups by linking together particles separated by less than 0.2 of the mean inter particle separation \citep{davis85}. The SUBFIND algorithm \citep{springel01} is then applied to each FOF group to split it into a set of self-bound sub halos. The central subhalo is defined as the most massive subunit of a FOF group. We construct subhalo merger trees which link each subhalo at each epoch to a unique descendent in the following epoch. These merger trees allow us to track the formation history of each (sub)halo that is identified at $z=0$. \cite{springel01} and \cite{bk09} provide a detailed description of these simulations and the post-processing techniques.

\subsection{Subhalo Abundance Matching}

Subhalo abundance matching (SHAM) is a technique for assigning galaxies to simulated dark matter halos and subhalos. The essential assumptions are that all galaxies reside in identifiable dark matter substructures and that luminosity or stellar mass of a galaxy is monotonically related to the potential well depth of its host halo or subhalo. Some implementations use the maximum of the circular velocity profile as the indicator of potential well depth, while others use halo or subhalo mass. The first clear formulations of SHAM as a systematic method appear in \cite{conroy06} and \cite{vale06}, but these build on a number of previous studies that either test the underpinnings of SHAM or implicitly assume SHAM-like galaxy assignment \citep[e.g.][]{colin99,kravtsov04,nagai05}.

N-body simulations produce subhalos that are located within the virial radius of halos. The present mass of subhalos is a product of mass built up during the period when the halo evolves in isolation and tidal mass loss after it enters the virial radius of a more massive halo \citep[e.g.][]{kravtsov04,kazantzidis04}. The stellar component, however, is at the bottom of the potential well and more tightly bound making it less likely to be affected by tidal forces. Therefore, several authors \citep[e.g.][]{conroy06,vale06} argue that the properties of the stellar component should be more strongly correlated with the subhalo mass at the epoch of accretion rather than at $z=0$.

\cite{vale06} apply a global statistical correction to subhalo masses relative to halo masses (as do \citealt{weinberg08}), while \cite{conroy06} explicitly identify subhalos at the epoch of accretion and use the maximum circular velocity at that epoch. Our formulation here is similar to that of \cite{conroy06}, though we use mass rather than circular velocity. Specifically, we assume a monotonic relationship between galaxy luminosity and halo mass at the epoch of accretion and determine the form of this relation by solving the implicit equation
\begin{equation}
 n_{\rm S}(>M_r) = n_{\rm H}(>M_{\rm H}),
\end{equation}
where $n_{\rm S}$ and $n_{\rm H}$ are the number densities of galaxies and halos, respectively, $M_r$ is the galaxy $r$-band magnitude threshold, and $M_H$ is the halo mass threshold chosen so that the number density of halos above it is equal to the number density of galaxies in the sample.
The quantity $M_{\rm H}$ is defined as follows:
\begin{equation}
M_{\rm H}=\begin{cases}
M_{\rm halo} (z=0) & \text{for distinct halos},\\
M_{\rm halo} (z=z_{\rm sat})& \text{for subhalos},
\end{cases}
\end{equation}
where z$_{\rm sat}$ is the epoch when a halo first enters the virial radius of a more massive halo. 

\cite{guo12} compare the mass functions of SHAM selected subhalos (see their figure 1) in MS and MS-II finding good convergence over the mass range where both simulations have adequate resolution. At $z=0$, they find that the (SHAM selected) subhalo mass function converges for subhalos with more than $\sim10^3$ particles at infall. This is more than an order of magnitude greater than the level at which the halo mass function converges \citep{bk09} because subhalos experience tidal stripping after infall, and must therefore have a larger number of particles at infall to be reliably identified at $z=0$. 

\begin{figure}
\centerline{
\epsfxsize=84mm
\epsfbox{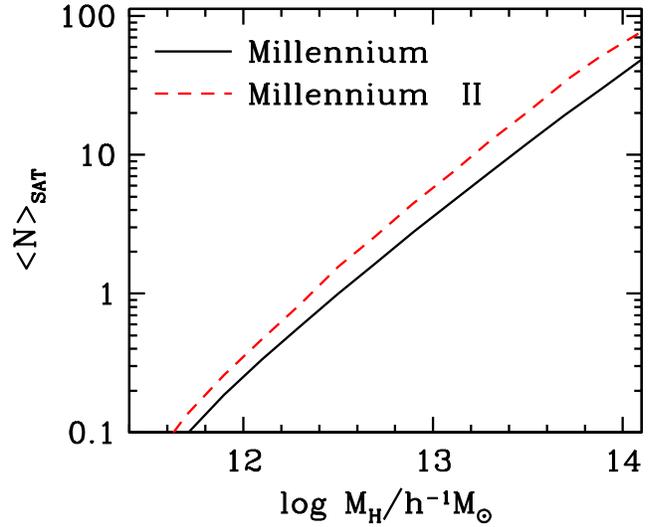}
}
\caption{
Mean number of SHAM selected subhalos per halo  as a function of parent halo 
mass at $z=0$. Subhalos above an infall mass threshold of 8.6$\times$10$^{10}$ 
$h^{-1}M_{\odot}$ are selected which 
corresponds to 100 times the particle mass in the Millennium Simulation.
}
\label{fig:sub1}
\end{figure}

Fig. \ref{fig:sub1} shows the satellite occupation function as a function of halo mass i.e. the mean number of SHAM selected satellite subhalos per parent halo as a function of parent 
halo mass. In both MS and MS-II, we rank order subhalos identified at $z=0$ by $M_{\rm H}$, mass at $z=0$ for central 
subhalos and at infall for satellite subhalos. We then select subhalos above a mass threshold of 8.6$\times$10$^{10}$ 
$h^{-1}$M$_{\odot}$ which is 100 times the particle mass in MS. Because of its higher resolution (factor of $\sim$ 100 difference in 
particle mass), more satellite subhalos are identified in MS-II compared to MS across the range of parent halo masses. 

In MS, there is a missing population of satellite subhalos that have fewer than $10^3$ particles at infall and cannot be 
identified at $z=0$ because of subsequent stripping, although similar objects are identified in the higher resolution MS-II. In order to exploit the larger box size and better statistics in MS, we construct a subhalo catalogue, Millennium Simulation Plus (MS+), that matches the satellite occupation function of MS-II. To achieve this, we augment the subhalos identified at $z=0$ in MS with additional subhalos that are identified at high $z$, but cannot be identified at $z=0$. These subhalos are randomly chosen from among disrupted subhalos that were above the mass threshold at infall. The number of additional subhalos in each parent halo mass bin is set so as to obtain a match to the MS-II satellite occupation function. The $z=0$ positions of each disrupted subhalo is obtained by tracking the position of its most bound particle from $z_{sat}$ to $z=0$.

The difference in the satellite occupation functions between MS and MS-II depends on the mass threshold that is applied, and for a mass threshold above $\sim10^{12} h^{-1}$M$_{\odot}$, they converge. Therefore, the MS+ catalogue has to be freshly constructed for each mass threshold that is required. While fitting the SDSS data, we carry out the procedure to construct MS+ with a mass threshold which gives the galaxy number density of the observed sample.

\begin{figure}
\centerline{
\epsfxsize=84mm
\epsfbox{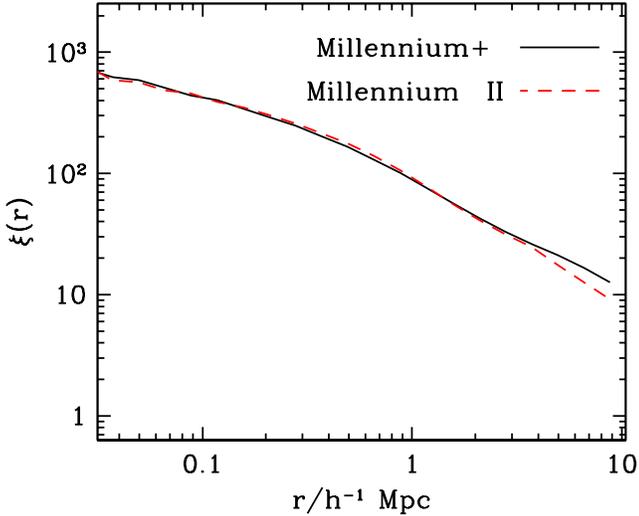}
}
\caption{
Two-point correlation function of $z=0$ subhalos that are 
above an infall mass threshold. In the Millennium II simulation, we only include
subhalos identified at $z=0$, while in  the 
Millennium simulation, we add a fraction of disrupted subhalos to those 
identified at $z=0$ so that we match the mean number of subhalos per 
parent halo as a function of parent halo mass in Millennium II (see Fig. 
\ref{fig:sub1}). 
}
\label{fig:sub2}
\end{figure}

Fig. \ref{fig:sub2} shows the two-point correlation function of SHAM selected subhalos in MS-II compared to MS+. On scales not affected by box size effects, the auto correlation functions extracted from the MS-II simulation and our MS+ method described above show strong agreement demonstrating the efficacy of this procedure.

\subsection{Rescaling Technique}

Using the clustering of galaxies at $z=0$ to infer cosmological parameters requires modelling the growth of density perturbations in the non-linear regime. In principle, one would need a numerical simulation for each set of cosmological parameters. However, running such a suite of high resolution cosmological N-body simulations to
adequately sample the cosmological parameter space is computationally expensive.

Instead, we rescale our simulation from its ``native" cosmology to different cosmologies with slightly different parameters, using a technique that involves rescaling the box length, velocity and mass units and relabelling output times. The first attempt to rescale the simulation output to a different cosmology was carried out by \cite{zheng02}. \cite{harker07} used this rescaling technique in combination with semi-analytic models of galaxy formation to constrain the cosmological parameters. More recently, \cite{angulo10} presented the technique of rescaling the output of a cosmological simulation from its ``native" cosmology to a cosmology with different parameters as a systematic algorithm. 
 
In this paper, we use the \cite{angulo10} algorithm to generate (sub)halo catalogues for a given set of cosmological parameters. Following \cite{ruiz11}, we do not apply the final step of the \cite{angulo10} algorithm which involves adjusting the amplitudes of large-scale linear modes which would in any case have negligible effects on the clustering statistics considered in this paper.

\cite{angulo10} and \cite{ruiz11} find good agreement between the halo catalogues extracted from simulations scaled to a target cosmology and simulations run with the target cosmology.  \cite {guo12} extend this comparison to subhalos finding good agreement between subhalo mass functions extracted from rescaled and ``native" simulations.

\section{Cosmological Constraints From The Galaxy Two-Point Correlation Function}

We probe the cosmology by fitting the observed galaxy two-point correlation function of a volume limited sample of galaxies. To generate a theoretical prediction for the two-point correlation function for a given cosmology, we implement the following steps. First, we rescale our simulation to the target cosmology using the procedure described in \S2.4. We then select (sub)halos using the SHAM procedure described in \S2.3. Briefly, subhalos above a mass threshold at infall for satellite subhalos, and at $z=0$ for central subhalos are selected. Our subhalo mass threshold is set so that the number density of SHAM selected subhalos in our simulation is equal to the number density of observed galaxies in the sample we are fitting to (see equation 2). Finally, we compute the projected two-point correlation function of these SHAM selected subhalos in the simulation cube.

\begin{figure}
\centerline{
\epsfxsize=84mm
\epsfbox{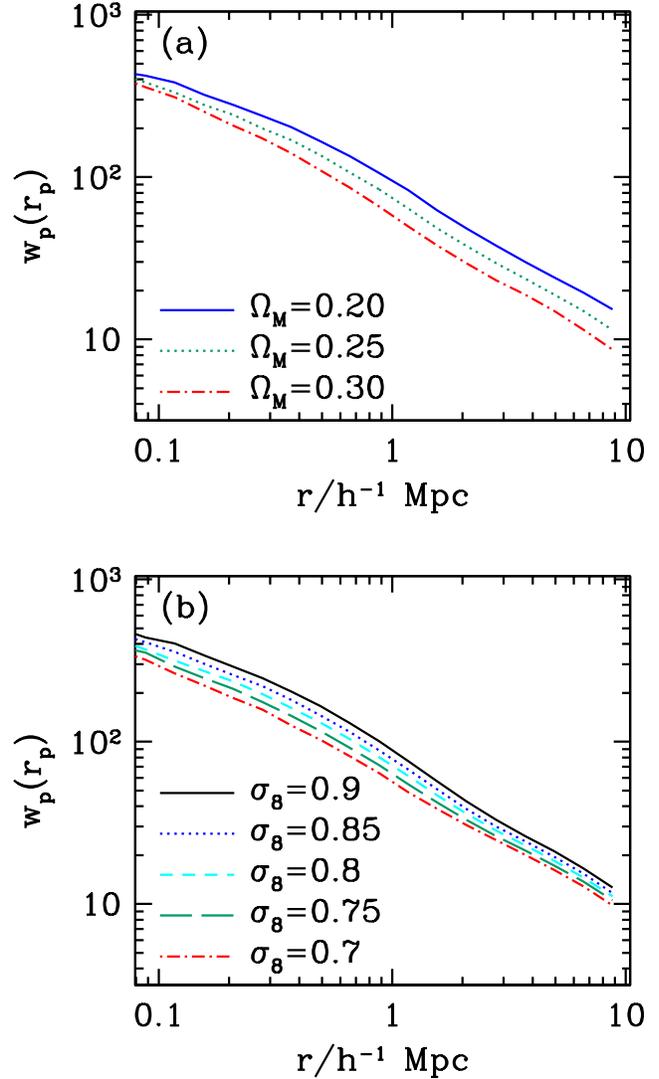}
}
\caption{
A pedagogical demonstration of the method employed in this paper. The effect on the two-point correlation function of SHAM selected
subhalos of changing $\Omega_{\rm M}$ at fixed $\sigma_8$ in panel (a), and of 
changing $\sigma_8$ at fixed $\Omega_{\rm M}$ in panel (b).
}
\label{fig:sub3}
\end{figure}

Fig. \ref{fig:sub3} provides a pedagogical demonstration of our technique demonstrating that, in principle, it could be used to obtain cosmological constraints. Panel (a) shows the effect of changing $\Omega_{\rm M}$ on the two-point galaxy correlation function in a standard $\Lambda$CDM cosmology with all other cosmological parameters held constant. Decreasing $\Omega_{\rm M}$ boosts the two-point correlation function on all scales (see figure 8 of \citealt{zheng02} for a similar result). Panel (b) shows the effect of changing $\sigma_8$ on the two-point galaxy correlation function in a standard $\Lambda$CDM cosmology with all other cosmological parameters held constant. Increasing $\sigma_8$ boosts the two-point correlation function on all scales.

\section{Cosmological Constraints}

We determine constraints on the cosmological parameters by comparing the predicted two-point galaxy correlation function for a grid of models in the $\sigma_8$-$\Omega_{\rm M}$ plane to the SDSS observed two-point galaxy correlation function of galaxies brighter than $M_r$ = -18. We restrict our analysis to scales below 10 $h^{-1}$Mpc.

We compute the $\chi^2$ for each model using the full covariance matrix. The covariance matrix is calculated by \cite{zehavi10} using jackknife resampling (see \S2.2 of \citealt{zehavi10} for further details of the method). There are two free cosmological parameters, $\Omega_{\rm M}$ and $\sigma_8$, and no other free parameters in our model. 

We impose the following flat priors on our cosmological parameters such that 0.2 $\le$ $\Omega_{\rm M}$ $\le$ 0.35 and  
0.65 $\le$ $\sigma_8$ $\le$ 1. Rescaling our simulation to a model with a different amplitude of clustering relies on 
relabelling the simulation output epochs. Since our simulations are only run to $z=0$, we are restricted to $\sigma_8$ $\le$ 0.9. For 0.6 $\le$  $\sigma_8$ $\le$ 0.9, linearly interpolating log($\xi$) with $\sigma_8$ works accurately, and so we 
extend this scaling to extrapolate the galaxy two-point correlation function to generate predictions for models with $
\sigma_8$ $\ge$ 0.9. 

\begin{figure}
\centerline{
\epsfxsize=84mm
\epsfbox{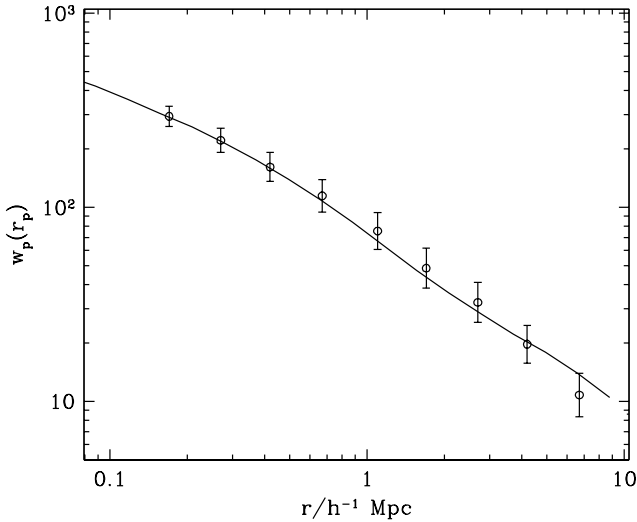}
}
\caption{
The solid curve is the galaxy two-point correlation function of our best-fit model with $\Omega_{\rm M}$ = 0.275 and $\sigma_8$ = 0.86. The points with error bars are the SDSS observed galaxy two-point correlation function from a volume limited sample of galaxies with $M_r$ $\le$ -18.0.
}
\label{fig:sub6}
\end{figure}

Fig. \ref{fig:sub6} shows the galaxy two-point correlation function for our best-fit model with $\Omega_{\rm M}$ = 0.275 and $\sigma_8$ = 0.86 plotted against the data. The $\chi^2$ for our best-fit model is 9.21. As we have two free parameters and nine data points, this corresponds to a $\chi^2$ per degree of freedom of 1.3. The probability of drawing a value $\ge$ 9.21 from a $\chi^2$ distribution with 7 degrees of freedom is 0.24, and so our best-fit model is an acceptable fit to the data.

\begin{figure}
\centerline{
\epsfxsize=84mm
\epsfbox{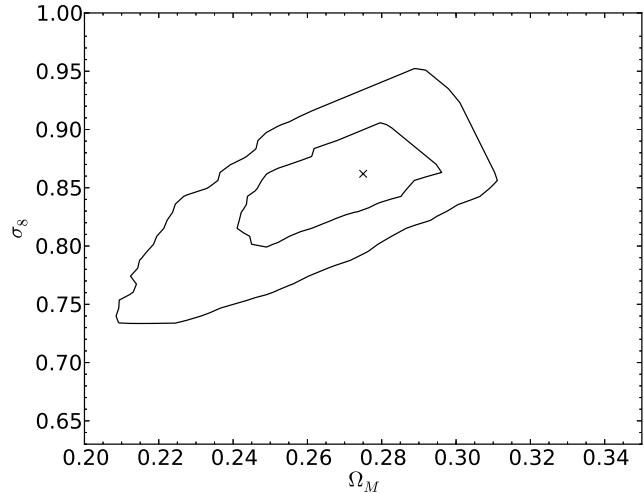}
}
\caption{
Joint constraint in the $\sigma_8$ - $\Omega_{\rm M}$ plane. The inner contour shows the boundary of the 68\% confidence region and the outer contour shows the 95\% confidence region. 
}
\label{fig:sub7}
\end{figure}

Fig. \ref{fig:sub7} shows our main result, cosmological constraints in the $\Omega_{\rm M}$-$\sigma_8$ plane. Because of the way $\Omega_{\rm M}$ and $\sigma_8$ affect the two-point galaxy correlation function (see \S3), our constraints on them are correlated. Note that only the portion of the outer contour with $\sigma_8$ $\ge$ 0.9 depends on our extrapolation of the two-point correlation function. 

\begin{figure}
\centerline{
\epsfxsize=84mm
\epsfbox{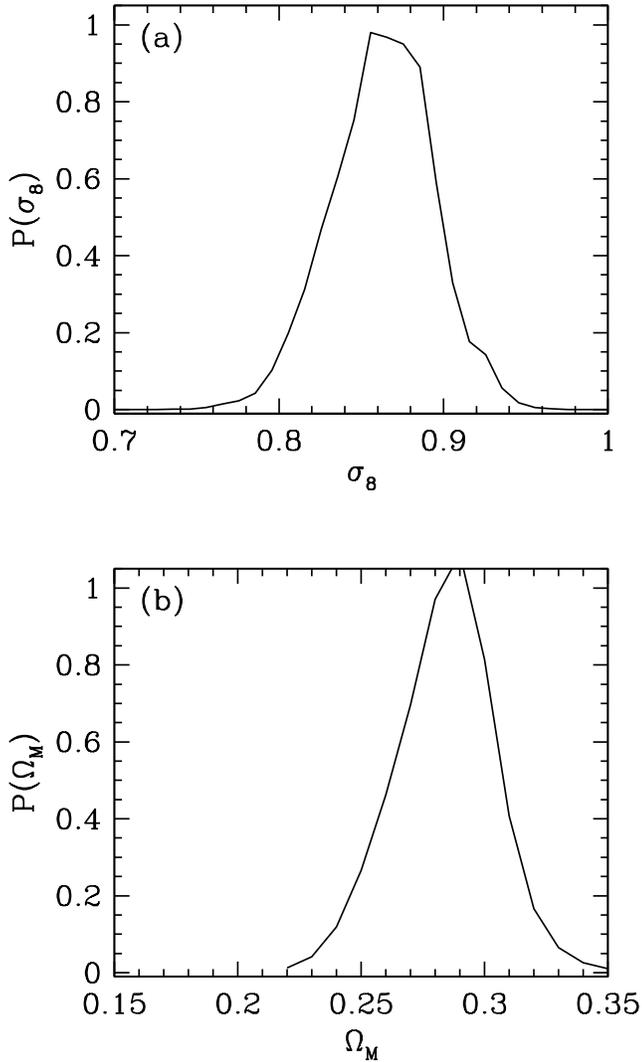}
}
\caption{
Marginalised posterior probability distribution of $\sigma_8$ in panel (a), and 
of $\Omega_{\rm M}$ in panel (b) from fitting the two-point galaxy correlation function.
}
\label{fig:sub8}
\end{figure}

Fig. \ref{fig:sub8} shows the marginalised probability distributions of $\Omega_{\rm M}$ and $\sigma_8$. Our constraints on the individual parameters are $\Omega_{\rm M}$ = 0.29 $\pm$0.03 and $\sigma_8$ = 0.86 $\pm$ 0.04 (68\%). 

We emphasize that our constraints on these parameters are obtained for fixed values of other cosmological parameters. We defer discussion of the implications of this to the final section. 

\section{Systematic Uncertainties}

We discuss the systematic uncertainties in our analysis, focusing on the uncertainties in our theoretical modelling. See \cite{zehavi10} for a discussion of observational uncertainties.

\subsection{Scatter in SHAM}

SHAM assumes a strictly monotonic relation between (sub)halo mass and galaxy $r$-band luminosity with zero scatter. While the assumption of zero scatter is idealised, several observational studies \citep[e.g.][]{zheng07,vdb07} indicate that the scatter in luminosity at fixed (sub)halo mass is small. While from the theory side, \cite{simha12} find a strong correlation between $r$-band luminosity and subhalo mass in their SPH simulations, with a small scatter of 0.15 dex in luminosity at fixed subhalo mass.

\begin{figure}
\centerline{
\epsfxsize=84mm
\epsfbox{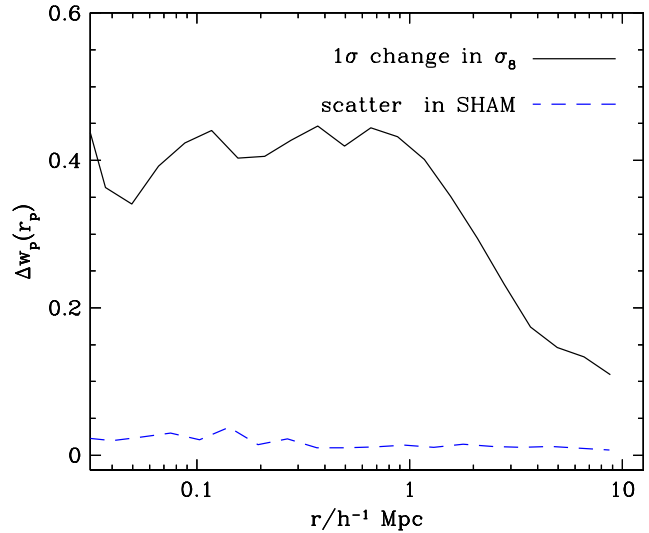}
}
\caption{
Effect of scatter in the subhalo mass - luminosity relation on the galaxy two-point correlation function compared to the effect of a 1$\sigma$ difference in $\sigma_8$. Fractional difference in the two-point galaxy correlation function compared to the fiducial model is plotted against length scale.
}
\label{fig:sys2}
\end{figure}

Fig. \ref{fig:sys2} shows the effect of scatter in the (sub)halo mass - luminosity relation on the predicted galaxy two-point correlation function. We use the results of \cite{simha12} to obtain an estimate of the scatter in subhalo mass in bins of galaxy luminosity, and perturb our subhalo masses by a random number drawn from a distribution centered on this value. We then compute the two-point correlation function for this sample of subhalos. We plot the fractional difference between the two-point correlation function of this sample and the fiducial model that assumes no scatter between (sub)halo mass and luminosity. For comparison, we show the fractional difference in correlation functions between our fiducial model and a 1$\sigma$ change in $\sigma_8$ calculated by taking the mean of the absolute value of the fractional difference in the correlation function from a positive and negative 1$\sigma$ change in $\sigma_8$ from our best-fit value of $\sigma_8$. The effect of random scatter in the subhalo mass - luminosity relation on the galaxy two-point correlation function is negligible compared to the effect of changing $\sigma_8$ by one standard deviation.

Fig. \ref{fig:sys2} shows the effect of scatter in the subhalo mass - luminosity relation for a particular set of cosmological parameters and a particular realisation of the scatter. However, our investigations lead us to believe that the effect of scatter in the subhalo mass - luminosity relation is not sensitive to small changes in the cosmological parameters.

Random scatter in the subhalo mass - luminosity relation has negligible impact on the galaxy/subhalo two-point correlation function because the effect of such scatter is only seen at the boundary of our sample. Incorrect identification of a galaxy's host halo does not affect our result as long as the halo hosts a galaxy that is sufficiently luminous to be included in our sample. However, this effect could potentially be a significant source of uncertainty for other statistics that depend on correctly identifying the host halos of galaxies.

\subsection{Satellite Galaxy Fraction}

SHAM relies on accurately identifying substructures and recovering their properties. Since we trace the $z_{\rm{sat}}$ progenitors of $z=0$ substructures, our results are unlikely to be affected by random fluctuations in the density field of halos that may be spuriously identified as substructures. However, if subhalos hosting satellite galaxies that have merged with the central galaxy of the halo are identified as substructures, we would overestimate the halo occupation of massive halos. Conversely, subhalos that fall into more massive halos and lose a substantial fraction of their mass due to tidal stripping may no longer be resolved in the simulation at $z=0$ although they may still host satellite galaxies.

\cite{simha12} find that for a fixed subhalo mass, satellite galaxies in their SPH simulation are typically somewhat less luminous than central galaxies because of differences in the ages of their respective stellar populations. If this were to be true of the real Universe, SHAM would overestimate the satellite galaxy fraction.

\begin{figure}
\centerline{
\epsfxsize=84mm
\epsfbox{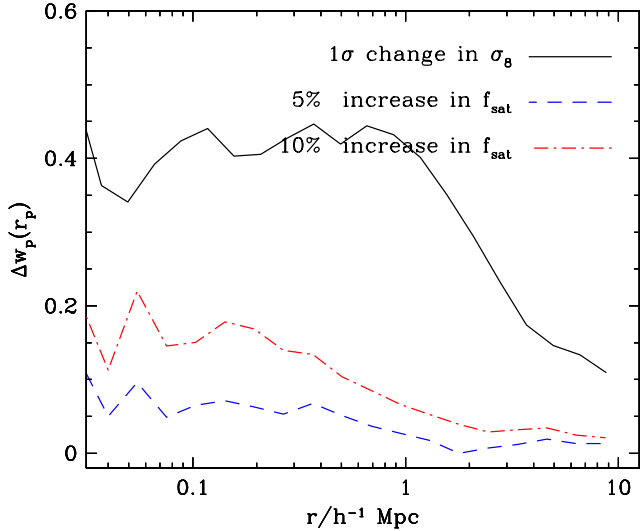}
}
\caption{
Effect of changing the fraction of satellite galaxies on the galaxy two-point correlation function compared to the effect of a 1$\sigma$ difference in $\sigma_8$. Fractional difference in the two-point galaxy correlation function compared to the fiducial model is plotted against length scale.
}
\label{fig:sys1}
\end{figure}

Fig. \ref{fig:sys1} shows the effect of changing the satellite galaxy fraction. To increase the satellite galaxy fraction, we include additional ``merged" subhalos i.e. subhalos that exist at high redshift, but have merged by $z=0$ using its most bound particle when it was last identified to track its position to $z=0$. We then compute the two-point correlation function for this sample of subhalos. We plot the fractional difference between the two-point correlation function of this sample and the fiducial model. For comparison, we again show the fractional difference in correlation functions between our fiducial model and a 1$\sigma$ change in $\sigma_8$. 

Increasing the fraction of satellite galaxies increases the number of close pairs and thus boosts the two-point correlation function on small scales. A small increase is also seen on large scales because highly biased objects in massive halos are present in a larger number of pairs compared to a sample with fewer satellite galaxies. 

An error of 10\% in the satellite galaxy fraction would alter our best-fit estimate of 
$\sigma_8$ by roughly half a standard deviation making this a more significant source of systematic uncertainty than scatter in the (sub)halo mass - luminosity relation.  


\subsection{Rescaling Technique}

\cite{ruiz11} compare the properties of halos in a rescaled simulation to a simulation run with the target cosmological parameters. They find that over 99\% of halos with more than 50 particles are recovered in the rescaled simulation. In the rescaled simulations, the masses of halos are systematically underestimated by $\sim$5\%. But because we use halo mass only to assign a rank to subhalos, we are not affected by this systemic bias. However, it would be a source of error if a statistic that depended directly on halo mass were being used. The rescaled halo mass - ``native" halo mass relation displays scatter. For our purpose, it would mimic the effect of scatter in the subhalo mass - luminosity relation discussed earlier. The positions of halos in their rescaled simulations are recovered to a precision of greater 100 $h^{-1}$kpc.

\subsection{Other Galaxy Samples}

Our cosmological constraints have been obtained by fitting our model to the observed clustering of one volume-limited sample of galaxies with $M_r$ $\le$ -18. As a consistency check, we compare the clustering predictions of our model with four other volume limited samples of galaxies taken from SDSS \citep{zehavi10} with $M_r$ $\le$ -18.5, -19, -19.5 and -20.5 corresponding to a number density of galaxies of 2.311, 1.676, 1.12 and 0.318 10$^{-2}$$h^{-3}$Mpc$^3$ respectively. The sample with $M_r$ $\le$ -21.5 corresponds approximately to galaxies brighter than $L_*$, the characteristic galaxy luminosity above which the number density of galaxies falls exponentially.

\begin{figure}
\centerline{
\epsfxsize=84mm
\epsfbox{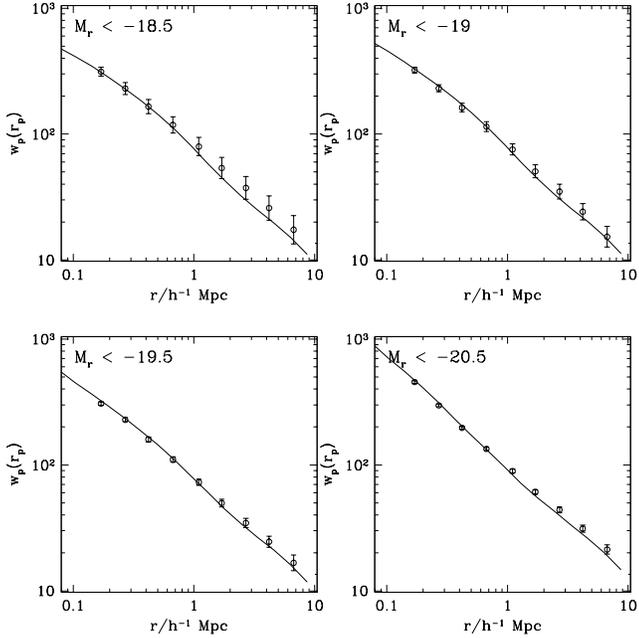}
}
\caption{
In each panel, the points with error bars are the SDSS observed galaxy two-point correlation function in a volume limited sample of galaxies brighter than $M_r$ = -18.5, -19, -19.5 and -20.5. The solid curve in each panel is the galaxy two-point correlation function predicted by our best-fit model with $\Omega_{\rm M}$ = 0.275 and $\sigma_8$ = 0.86 for the corresponding galaxy sample. 
}
\label{fig:sub9}
\end{figure}

For each of these samples, we generate the predicted galaxy two-point correlation function for our best-fit cosmological model by measuring the clustering of SHAM selected subhalos above an infall mass threshold determined so that the number density of subhalos is equal to the number density of galaxies in the sample. Each panel of Fig. \ref{fig:sub9} shows the galaxy two-point correlation function predicted by our best-fit model plotted against the observed galaxy two-point correlation function for the corresponding sample. Formally, our model is a good fit to the data for the volume limited samples with $M_r$ $\ge$ -18.5, -19 and -19.5, but not for the brightest sample with $M_r$ $\ge$ -20.5.

The error bars in each panel show the diagonal errors on the observed correlation function. For our main sample and samples with relatively high number density, the error on the correlation function extracted from our simulation is negligible compared to the observational errors, and can therefore be ignored. However, because of our finite simulation volume, as we go to brighter samples with low number density, the errors on our predicted correlation function are sufficiently large to be comparable to the observational errors. 

When we fit models to these other galaxy samples by allowing the cosmological parameters to vary, we find that for each sample, the best-fit cosmological parameters are slightly different, but consistent with the constraints from our main sample ($M_r$ $\le$ -18). In principle, tighter constraints could be obtained from a joint fit to all the datasets, but this would require an estimate of their covariance.

\section{Complementary Constraints on Cosmology}

As we discussed earlier, (see Fig. \ref{fig:sub7}), our constraints on $\Omega_{\rm M}$ and $\sigma_8$ are correlated because of the way the changes in these parameters affect the galaxy auto-correlation function(see Fig. \ref{fig:sub3}). To a certain extent, the effect of a higher $\Omega_{\rm M}$ can be compensated for by a higher $\sigma_8$ if only the galaxy two-point correlation function is used to constrain the cosmology. But complementary constraints on the cosmology can be obtained by probing other observables for which predictions can be generated using the same SHAM model used in this paper.

\begin{figure}
\centerline{
\epsfxsize=84mm
\epsfbox{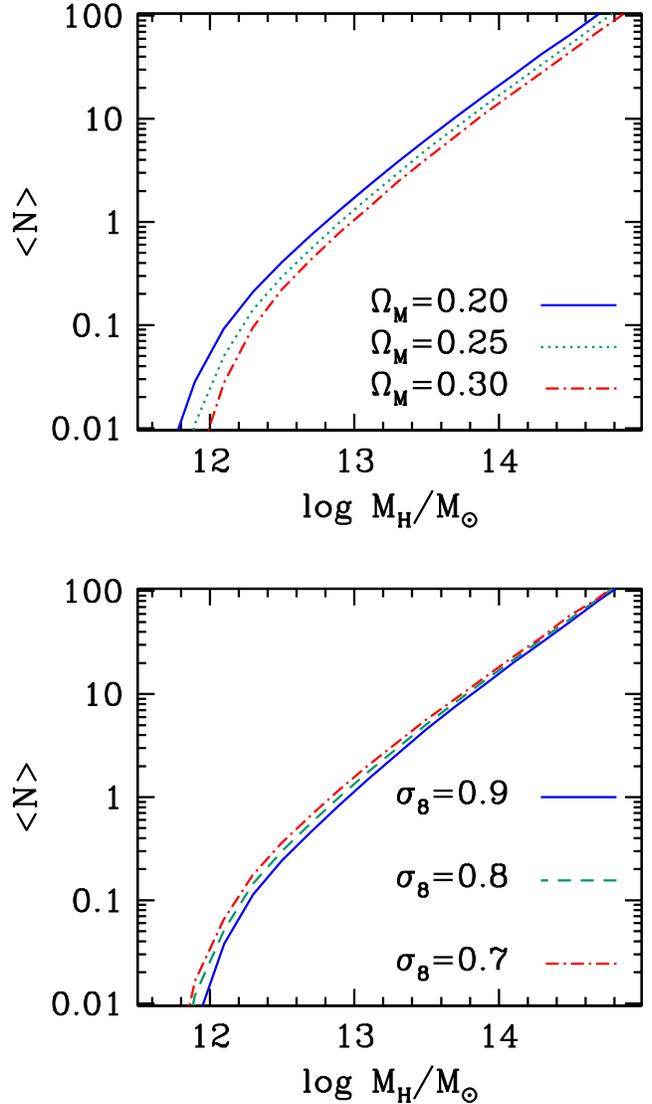}
}
\caption{
Panel (a) shows the effect of changing $\Omega_{\rm M}$ at fixed $\sigma_8$, and panel (b) shows the 
effect of changing $\sigma_8$ at fixed $\Omega_{\rm M}$ on the mean number of satellite galaxies per halo as a function of halo mass.
}
\label{fig:sub10}
\end{figure}

Fig. \ref{fig:sub10} shows the effect of the cosmological parameters on the mean number of satellite galaxies per halo, ${\langle N \rangle}_{\rm sat}$ as a function of halo mass in our simulation using our rescaling technique and SHAM model. The number of satellite galaxies in a given halo, $N_{\rm sat}$ = $N$ - 1 where $N$ is the number of galaxies in the halo. Panel (a) shows the effect of changing $\Omega_{\rm M}$ at fixed $\sigma_8$. Changing $\Omega_{\rm M}$ affects the halo mass function making halos with a given number of satellites more massive for higher $\Omega_{\rm M}$. Panel (b) shows the effect of changing $\sigma_8$ at fixed $\Omega_{\rm M}$. Increasing $\sigma_8$ increases the number of halos of a given mass. Therefore, for a fixed number density of galaxies, there must be fewer galaxies in each halo compared to models with lower $\sigma_8$. 

Decreasing $\sigma_8$ and decreasing $\Omega_{\rm M}$, both boost ${\langle N \rangle}_{\rm sat}$ rather than counteracting each other as they do for the two-point correlation function. Moving along the degeneracy curve in the $\Omega_{\rm M}$-$\sigma_8$ plane in Fig. \ref{fig:sub7} would generate different and easily distinguishable distributions of ${\langle N \rangle}_{\rm sat}$. Therefore, for a given cosmology, the additional requirement of matching the galaxy two-point correlation function strongly constrains the HOD, and consequently a direct measure of the HOD would place constraints on the cosmology that are complementary to our constraints from fitting the two-point correlation function. 

One possible direct measure of the HOD is the mean number of galaxies in halos of a given mass. The ratio of these quantities, $M/N$, is used by \cite{tinker11} to place constraints on cosmological parameters. 



\section{Discussion and Conclusions}

We have placed constraints on $\sigma_8$ and $\Omega_{\rm M}$ by comparing the SDSS observed projected galaxy two-point correlation function for a volume-limited sample of galaxies with $M_r$ $\le$ -18 to our model predictions generated using N-body simulations rescaled to the target cosmology using the technique of \cite{angulo10} and populated with galaxies using subhalo abundance matching (SHAM). 

Assuming a flat $\Lambda$CDM cosmology with $n_S$=1, we find $\Omega_{\rm M}$ = 0.29 $\pm$0.03 and $\sigma_8$ = 
0.86 $\pm$ 0.04 at 68\% confidence. 

\begin{figure}
\centerline{
\epsfxsize=84mm
\epsfbox{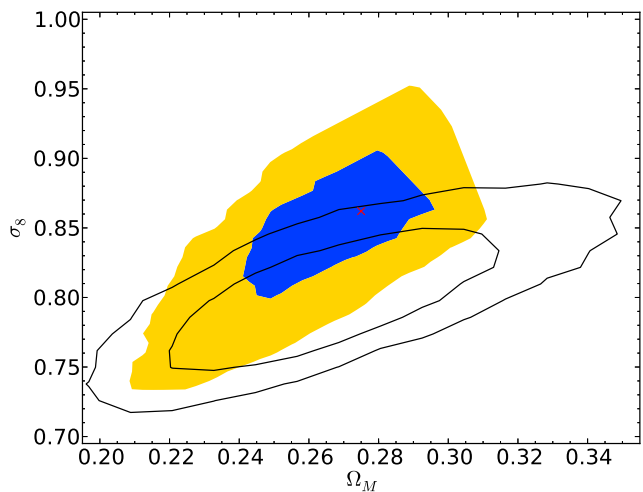}
}
\caption{
Joint constraint in the $\sigma_8$ - $\Omega_{\rm M}$ plane. The inner contour shows the boundary of the 68\% confidence region and the outer contour shows the 95\% confidence region. The filled contour is the result from this work while the black solid open contours are from WMAP7 \citep{komatsu11}. 
}
\label{fig:sub11}
\end{figure}

Fig. \ref{fig:sub11} compares our constraint in the $\Omega_{\rm M}$-$\sigma_8$ plane to constraints from WMAP7 \citep{komatsu11}. Our estimates of  both $\sigma_8$ and $\Omega_{\rm M}$ are high compared to WMAP7, but are consistent at the $\sim$2 $\sigma$ level. 

Our constraints are obtained for fixed values of other cosmological parameters. In contrast, the WMAP7 constraints are obtained by marginalizing over all other cosmological parameters. One significant implication of this is in regard to the shape of the power spectrum. We have assumed that the value of the primordial spectral index, $n_S$=1 in our simulation. In contrast, the WMAP-7 best-fit value of $n_S$ is 0.96, and $n_S$=1 is excluded at more than 2 $\sigma$. While we are unable to comment on the effect of setting  $n_S$ = 0.96 on our constraints, forcing $n_S$ = 1 while fitting the CMB data would result in a higher best-fit value of $\sigma_8$. 

Our results are consistent with and comparable to \cite{tinker11} who fit the SDSS galaxy two-point correlation function and $M/N$ (cluster mass to number ratio) using their HOD models, finding $\sigma_8$ = 0.85 $\pm$ 0.06 and $\Omega_{\rm M}$ = 0.29 $\pm$0.03. 

However there is some tension between our results and those of \cite{harker07} who use semi-analytic models to populate N-body simulations rescaled to a given cosmology using a technique similar to that of this paper, and fit to the SDSS clustering finding $\sigma_8$ = 0.97 $\pm$ 0.06. In contrast to \cite{harker07} who use a semi-analytic model of galaxy formation to populate their N-body simulation with galaxies, we use subhalo abundance matching which only assumes a monotonic relationship between galaxy luminosity and subhalo mass at infall. Secondly, the resolution of MS-II is $\sim$ 2000 times higher than the simulations used by \cite{harker07}. Thirdly, \cite{harker07}  
use a Monte Carlo scheme to generate a merger tree for a halo based on its mass. Consequently, satellite galaxy positions are not obtained from the simulation, and they are instead placed on random particles within the halo. We are unable to quantify the effects of each of these factors. However, repeating the work of \cite{harker07} with a high resolution N-body simulation and merger trees and subhalo positions extracted from the N-body simulation would be interesting and could potentially reveal the source of the tension.

 Besides the CMB and galaxy clustering data, several other methods have been employed to constrain $\sigma_8$. For example, \cite{mandelbaum12} find $\sigma_8$$(\Omega_{\rm M}/0.25)^{0.57}$ = 0.80 $\pm$ 0.05 using galaxy-galaxy lensing, \cite{seljak05} find $\sigma_8$ = 0.90 $\pm$ 0.03 by combining their analysis of the Lyman $\alpha$ forest power spectrum with CMB results. 

Because our simulation is only run to $z=0$, we are unable to rescale our subhalo catalogue to cosmologies with $\sigma_8$ $\ge$ 0.9. Our model predictions for $\sigma_8$ $\ge$ 0.9 are based on extrapolating the correlation function as a function of the cosmological parameters. While this appears to be a reasonable approximation, it is likely that it has shortcomings that will be exposed by a simulation that is run with a higher value of $\sigma_8$.

We have examined some potential sources of systematic error in \S3 finding that random scatter in the luminosity - halo mass relation does not affect our results significantly. However, if the scatter were to be correlated with other properties of the galaxy population, our results would be affected differently. We also examine the impact of a systematic error in the fraction of satellite galaxies which could either be underestimated or overestimated in the models either for numerical reasons relating to identification of subhalos or due to systematic differences in the growth of central and satellite galaxies that violate the implicit assumptions of SHAM. We find that a 10\% change in the fraction of satellite galaxies would alter our constraint on $\sigma_8$ by half a standard deviation. The two-point galaxy correlation function on small scales is sensitive to the distribution of galaxies within massive halos. Therefore, our results would be significantly affected if the distribution of galaxies in halos were to systematically differ from the distribution of SHAM selected subhalos in N-body simulations.








Despite these caveats, we emphasize that the technique presented in this paper can provide tight constraints on the cosmology using only low $z$ data. The remarkable tightness of our constraint arises from the fact that unlike statistical descriptions of the distribution of galaxies in halos provided by HOD models or the CLF (Conditional Luminosity Function), we do not have the freedom to define the HOD. Furthermore, our model does not make any assumptions about galaxy bias or the detailed physics of galaxy formation except for requiring a monotonic relationship between galaxy luminosity and subhalo mass at infall. Additionally, we assume that galaxies are the observational counterparts of subhalos identified at $z=0$ in an N-body simulation, and that each observed galaxy can be associated with a subhalo.

Although the potential for statistical tightening of the parameter constraints is limited, substantial improvements in the robustness of this technique can be achieved by future investigations. Firstly by adopting the shape of the power spectrum $P(k)$ inferred by the latest CMB observations. Secondly, a large volume simulation with the resolution of MII or higher would remove the need for the correction to the subhalo fraction that we apply in this work and the uncertainties associated with it. Systematic uncertainties associated with the rescaling technique scale with the magnitude of the difference in cosmological parameters between the ``native" and rescaled cosmology. To minimise this, it would be useful to carry out a suite of simulations with different cosmological parameters that span the requisite range, and only apply the rescaling technique to generate predictions for intermediate values of the parameters. Finally, further investigations of the SHAM technique, particularly with respect to scatter in the luminosity - halo mass relation, and the distribution of subhalos within halos will help to clarify potential sources of systematic errors involved in this technique.

\section * {ACKNOWLEDGEMENTS}
We are grateful to John Helly for technical assistance at various stages of this work. We thank Idit Zehavi for providing the covariance matrices. We thank Carlton Baugh for providing invaluable technical 
assistance, and Carlos Frenk and David Weinberg for useful discussions. The analyses presented in this paper used the
Cosmology Machine supercomputer at the ICC, which is part of the DiRAC Facility jointly funded by STFC, the Large 
Facilities Capital Fund of BIS, and Durham University.




\bibliographystyle{mn2e}

\end{document}